\documentclass[conference]{IEEEtran}
\IEEEoverridecommandlockouts
\usepackage{cite}
\usepackage{amsmath,amssymb,amsfonts}
\usepackage{graphicx}
\usepackage{textcomp}
\usepackage{xcolor}
\usepackage{makecell}
\usepackage{multirow}
\usepackage{balance}

\usepackage{array}

\def\BibTeX{{\rm B\kern-.05em{\sc i\kern-.025em b}\kern-.08em
    T\kern-.1667em\lower.7ex\hbox{E}\kern-.125emX}}
\begin{document}

\title{Multi-objective Optimal Control of\\
Dynamic \!Integrated \!Model \!of \!Climate \!and \!Economy:\\
Evolution in Action
}

\author{

    \IEEEauthorblockN{
        Mostapha Kalami Heris\textsuperscript{a,b}\\
        kalami@faradars.org\\
        {}
    }

    \IEEEauthorblockA{
        \textsuperscript{a}
        \textit{Intelligent Systems Lab.} \\
        \textit{Research and Development Department}\\
        FaraDars Online Education Organization\\
        Tehran, Iran
    }

    \and

    \IEEEauthorblockN{
        Shahryar Rahnamayan\textsuperscript{b}, SMIEEE\\
        shahryar.rahnamayan@uoit.ca\\
        {}
    }

    \IEEEauthorblockA{
        \textsuperscript{b}
        \textit{Nature-Inspired Computational Intelligence (NICI) Lab.} \\
        \textit{Department of Electrical, Computer and Software Engineering}\\
        Ontario Tech University\\
        Oshawa, Canada
    }

}

\maketitle

\begin{abstract}
One of the widely used models for studying economics of climate change is the Dynamic Integrated model of Climate and Economy (DICE), which has been developed by Professor William Nordhaus, one of the laureates of the 2018 Nobel Memorial Prize in Economic Sciences. Originally a single-objective optimal control problem has been defined on DICE dynamics, which is aimed to maximize the social welfare. In this paper, a bi-objective optimal control problem defined on DICE model, objectives of which are maximizing social welfare and minimizing the temperature deviation of atmosphere. This multi-objective optimal control problem solved using Non-Dominated Sorting Genetic Algorithm II (NSGA-II) also it is compared to previous works on single-objective version of the problem. The resulting Pareto front rediscovers the previous results and generalizes to a wide range of non-dominant solutions to minimize the global temperature deviation while optimizing the economic welfare. The previously used single-objective approach is unable to create such a variety of possibilities, hence, its offered solution is limited in vision and reachable performance. Beside this, resulting Pareto-optimal set reveals the fact that temperature deviation cannot go below a certain lower limit, unless we have significant technology advancement or positive change in global conditions.
\end{abstract}


\section{Introduction} \label{sec_introduction}
Global warming has been one of the most crucial problems, and maybe the greatest related to our real-world environment, which humankind is seriously faced with. As a result of industrialization and economic development, emission of greenhouse gases, specially carbon dioxide (CO\textsubscript{2}), caused a significant increase in the mean temperature of the atmosphere by 1 $^{\circ}$C to 1.5 $^{\circ}$C, according to international reports \cite{stocker_2013}, relative to pre-industrial era. This deviation of temperature, caused many changes in the global climate and environment, including but not limited to melting of ice lands \cite{solomon_2007}, increasing trend towards larger wildfires \cite{borunda_2018,barbero_2015}, increased heat content of deep level oceans \cite{kennedy_2010}, sea level rise \cite{cazenave_2014}, increased heavy precipitation \cite{wuebbles_2017}, and earlier timing of spring events \cite{solomon_2007,rosenzweig_2007}.

According to a special report \cite{masson_2018} from Inter-governmental Panel on Climate Change (IPCC), which is published in 2018, to keep the temperature deviation below 2 $^{\circ}$C, we will require to reach zero greenhouse gases emission by year 2050. Despite the fact that this target temperature is remaining in very same condition of present state, however, reaching this objective is very hard and it needs a global cooperation among all countries and nations. In order to prevent the destructive consequences of climate change, it must be taken care of every little bit of emission.

Scientists of the field use the Integrated Assessment Models (IAMs) to model the economy-climate interactions and quantify the damages of greenhouse emissions on the society and life on the planet. While many models proposed to describe the climate change economy, only three of them are mostly cited and commonly used as IAMs, all of which make possible estimation of Social Cost of Carbon (SCC). These models are as follows: Dynamic Integrated model of Climate and Economy (DICE) \cite{nordhaus_1975,nordhaus_1992,nordhaus_2007,nordhaus_2008,nordhaus_2010,nordhaus_2014,nordhaus_2016,nordhaus_2017}, Climate Framework for Uncertainty, Negotiation, and Distribution (FUND) \cite{anthoff_2013}, and Policy Analysis for the Greenhouse Effect (PAGE) \cite{hope_2012}. As an example, the United States Interagency Working Group (IWC) used these three models to estimate the social cost of carbon dioxide emissions.

These three models, DICE, FUND, and PAGE, are different. For example, while DICE is a general equilibrium model, as it models the economic growth as a state variable in the model, PAGE and FUND models are partial equilibrium models, hence, they do need to have an external input describing the economic factors. Also for FUND and PAGE models, there is not any optimization problems associated with the model, therefore, it is impossible to define an optimal control problem on these models. However, DICE is associated with an objective function and it can be expressed as an optimal control problem, accordingly.

In this paper, we are going to use DICE and define a multi-objective optimal control problem on this model. Perhaps, Professor William Nordhaus proposed first IAM model of the field in \cite{nordhaus_1975} and then developed that model to reach the DICE. First version of DICE is proposed by Nordhaus in 1992 \cite{nordhaus_1992}, and he improved and extended the model regularly in several years \cite{nordhaus_2007,nordhaus_2008,nordhaus_2010,nordhaus_2014,nordhaus_2016,nordhaus_2017}. Alongside the DICE model, there is a regional version of this model, named Regional Integrated model of Climate and Economy (RICE) \cite{nordhaus_2010}, which divides the whole world into 12 economic regions with their specific economic dynamics. In 2018, for his work in the field of climate change economy, Nordhaus received the Nobel Memorial Prize in Economic Sciences.
There are two versions of DICE model, which are mostly cited in the literature: DICE2013 \cite{nordhaus_2014} and DICE2016 \cite{nordhaus_2016,nordhaus_2017}. These two models have not any structural differences. But some of parameters and time window of these two editions of DICE model are different. In this paper, we will use the dynamics and parameters of the DICE2016 version.

Previously, Model Predictive Control (MPC) is used to solve the optimal control problem on DICE model in \cite{faulwasser_2018} and \cite{kellett_2019}. In these works, as well as original works of Nordhaus, the objective function is defined to be the Social Welfare, which incorporates both economic and environmental factors. However, to have a direct approach towards optimization of climate change and temperature deviation, while keeping optimizing the economic factors, we converted the problem into a multi-objective optimal control problem and used Non-Dominated Sorting Genetic Algorithm II (NSGA-II) to solve it. The results indicate that not only the result of MPC \cite{faulwasser_2018,kellett_2019} is discoverable by NSGA-II, but by having a complete set of Pareto-optimal solutions, we can make better decisions to reach lower temperature deviations levels.

This paper is organized as follows. In Section \ref{sec_dice_model} we will describe the dynamics and structure of DICE model in detail. In Section \ref{sec_problem}, we will define a multi-objective optimal control problem on DICE model, which we are going to solve it, in following sections. Section \ref{sec_algorithm} contains a brief description of NSGA-II, as well as the structure of solutions to the problem and the method used for generating initial population. In Section \ref{sec_results}, the results of the NSGA-II for the problem are presented and finally, Section \ref{sec_conclusion} concludes the paper.

\section{DICE Model} \label{sec_dice_model}
DICE model is a discrete-time model, and its latest version, starts from year 2015 as initial time. The time and discrete index are related to each other, according to the following equation:
\begin{equation}
    t = t_0 + (i-1) \cdot {\Delta}t \text{,}
    \label{eq_time}
\end{equation}
where $i$, $t$, $t_0$ and ${\Delta}t$ denote to discrete time index, time, starting year (2015), and sample time (5 years), respectively. We will use the discrete index in the equations of DICE model.

\subsection{Dynamics of Population} \label{sec_dice_model_pop}
Total population of the world is given by the following dynamic equation:
\begin{equation}
    L(i+1) = \left( \frac{1+L_a}{1+L(i)} \right)^{\ell_g} L(i) \text{,}
    \label{eq_pop_dynamics}
\end{equation}
where $L(i)$ denotes the global population at time step $i$, and $L_a$ and $\ell_g$ are parameters. Equation \eqref{eq_pop_dynamics} indicates that dynamics of global population is assumed to be independent from any other state variables of the system. The parameter $L_a$ is the maximum value of worldwide population $L(i)$. For current time, the exponential term in right-hand side of \eqref{eq_pop_dynamics} is greater than 1, because $L(i) \leqslant L_a$ for now. This means the population will increase by time. However, as global population increases, the exponential term will be closer and closer to 1. This will keep the population below $L_a$ and just in limit, $L(i)$ will reach it.

\subsection{Economic Dynamics} \label{sec_dice_model_economic}
According to Cobb-Douglas production function, the Gross Economic Output is expressed as:
\begin{equation}
    Y(i) = A(i) K(i)^\gamma L(i)^{1-\gamma} \text{,}
    \label{eq_gross_output}
\end{equation}
where $A(i)$, $K(i)$ and $\gamma$ denote technological efficiency (aka. total factor productivity), total capital, and capital elasticity constant, respectively.

Technological efficiency $A(i)$ is driven by the following dynamics:
\begin{equation}
    A(i+1) = \frac{A(i)}{1 - g_A \exp(-\delta_A (i-1){\Delta}t)} \raisebox{0.5ex}{\text{,}}
    \label{eq_tech_eff}
\end{equation}
where $g_A$ and $\delta_A$ are constant parameters. This equation expresses the dynamics of another state variable $A(i)$, which is assumed to be independent from other states in the DICE model. Denominator of the right-hand side of \eqref{eq_tech_eff} will converge to $1-g_A$ as $i \to \infty$. Therefore, in limit, the total progress of technological efficiency within a 5-year period, equals to $\frac{1}{1-g_A}$, which is approximately equal to 1.0823. This means, according to DICE model, technological efficiency improves about 8.23\% within 5 years, at most.

Total capital K(i)is given by the following dynamical equation:
\begin{equation}
    K(i+1) = (1-\delta_K)^{{\Delta}t}K(i) + I(i){\Delta}t \text{,}
    \label{eq_capital}
\end{equation}
where $\delta_K$ and $I(i)$ are respectively annual capital depreciation rate and total annual investment at time step $i$. Equation \eqref{eq_capital} says that, either capital is invested or remains as capital and depreciates.

The amount of investment is defined as:
\begin{equation}
    I(i) = s(i)Q(i) \text{,}
    \label{eq_investment}
\end{equation}
where $s(i) \in [0,1]$ is a control input, known as saving rate, and $Q(i)$ is Net Economic Output. After subtracting the investment from net economic output, the remaining amount will be total consumption $C(i)$, which is given by:
\begin{equation}
    C(i) = Q(i) - I(i) = [1-s(i)]Q(i) \text{.}
    \label{eq_consumption}
\end{equation}

Net economic output is defined as follows:
\begin{equation}
    Q(i) = [1-\Lambda(i)]\Omega(i)Y(i) \text{,}
    \label{eq_net_output}
\end{equation}
where $\Lambda(i)$ and $\Omega(i)$ are abatement cost fraction and climate damage function, respectively. These functions reduce the gross economic output $Y(i)$ to get the net output $Q(i)$ and they are defined by following relations:
\begin{equation}
    \Lambda(i) = \; \theta_1(i)\mu(i)^{\theta_2}
    \label{eq_abatement_cost}
\end{equation}
and
\begin{equation}
    \Omega(i) = \; \frac{1}{1 + \psi_1 T_\mathrm{AT}(i) + \psi_2 T_\mathrm{AT}(i)^2} \raisebox{0.5ex}{\text{,}}
    \label{eq_climate_damage}
\end{equation}
where $\mu(i) \in [0,1]$ is a control input, known as mitigation rate or abatement fraction, $\theta_1(i)$ is the cost of mitigation effort, $T_\mathrm{AT}(i)$ is deviation of atmosphere temperature with respect to year 1900, and $\theta_2$, $\psi_1$, and $\psi_2$ are parameters. Dynamics for $T_\mathrm{AT}(i)$ will be discussed in the next section. In damage function, $\psi_1$ is set to be zero and $\psi_2$ is set to have a total damage of 2 percent for $T_\mathrm{AT}(i) = 3$ $^{\circ}$C.

Also, the cost of mitigation effort $\theta_1(i)$ is given by the following equation:
\begin{equation}
    \theta_1(i) = \frac{p_b}{1000 \cdot \theta_2} (1-\delta_{pb})^{i-1} \sigma(i) \text{,}
    \label{eq_mitigation_effort}
\end{equation}
where $\sigma(i)$ is the global emission intensity of economic activity, which is defined in the next section, and $p_b$, $\delta_{pb}$, and $\theta_2$ are parameters.

The mitigation or abatement cost fraction $\Lambda(i)$ is related to emission intensity, hence, it connects the net economic output to carbon emission dynamics. On the other hand, the damage function $\Omega(i)$, is related to $T_\mathrm{AT}(i)$, i.e. atmosphere temperature deviation. This relates the economic dynamics to climate and carbon emissions in advance. Actually, \eqref{eq_net_output} is the part of model which relates the socio-economic parts of the model to the environment and climate.

\subsection{Carbon Dynamics} \label{sec_dice_model_carbon}
Global emission intensity dynamics is governed by:
\begin{equation}
    \sigma(i+1) = \frac{\sigma(i)}{\exp \big( {\Delta}t \cdot g_\sigma (1-\delta_\sigma)^{(i-1){\Delta}t} \big)} \raisebox{0.5ex}{\text{,}}
    \label{eq_emission_intensity}
\end{equation}
where $g_\sigma$ and $\delta_\sigma$ are parameters. This equation indicates that emission intensity only depends on time and it is not related to other state variables of the mode. The equilibrium point of \eqref{eq_emission_intensity} is $\sigma(\infty) = 0$. Because denominator of the right-hand side of the equation is always greater than 1. Hence, the value of emission intensity will be always decreasing, until it reaches zero in limit.

Total emission is the sum of emissions due to economic activity and emissions related to the change of land use and it is given by:
\begin{equation}
    E(i) = \sigma(i)[1-\mu(i)] Y(i) + E_\mathrm{Land}(i) \text{,}
    \label{eq_total_emission}
\end{equation}
where $E(i)$ and $E_\mathrm{Land}(i)$ are total emissions and land-use-related emissions, respectively. Equation \eqref{eq_total_emission} says that per one unit of gross economic output $Y(i)$, total emissions increase by $\sigma(i)$ units. Here, the mitigation control input $\mu(i)$ affects the emissions, and in case of full mitigation, total amount of economy related emissions can be cancelled out.

Emissions related to the change of land use is given by:
\begin{equation}
    E_\mathrm{Land}(i) = (1-\delta_{\mathrm{EL}}) E_\mathrm{Land}(i-1) = E_{\mathrm{L},0} (1-\delta_\mathrm{EL})^{i-1} \text{,}
    \label{eq_land_use_emissions}
\end{equation}
where $E_{\mathrm{L},0}$ is the estimation of initial emission due to land use in base year (2015), and $\delta_\mathrm{EL}$ is respective reduction rate. This equation indicates that land-use-related emission only depends on time and it is reducing as time passes. In this model, the total possible amount of emissions related to land use, from now to far future, is assumed to be constant, and according to \eqref{eq_land_use_emissions} it can be calculated as $(1 + \frac{1}{\delta_\mathrm{EL}}) E_{\mathrm{L},0}$.

In order to model the mass of carbon on the planet, three state variables introduced: the average mass of carbon in atmosphere, upper ocean carbon mass, and lower (deep) ocean carbon mass, which are denoted by $M_\mathrm{AT}$, $M_\mathrm{UP}$, $M_\mathrm{LO}$, respectively. Dynamics of these quantities are given by:
\begin{equation}
    \begin{bmatrix}
      M_\mathrm{AT}(i+1) \\
      M_\mathrm{UP}(i+1) \\
      M_\mathrm{LO}(i+1) \\
    \end{bmatrix}
    =
    \begin{bmatrix}
      \zeta_{11} & \zeta_{12} & 0 \\
      \zeta_{21} & \zeta_{22} & \zeta_{23} \\
      0 & \zeta_{32} & \zeta_{33} \\
    \end{bmatrix}
    \begin{bmatrix}
      M_\mathrm{AT}(i) \\
      M_\mathrm{UP}(i) \\
      M_\mathrm{LO}(i) \\
    \end{bmatrix}
    +
    \begin{bmatrix}
      \xi_2 \\
      0 \\
      0 \\
    \end{bmatrix}
    E(i) \text{,}
    \label{eq_carbon_dynamics}
\end{equation}
where the elements of state transition and input matrices of this state-space model are constant parameters.

According to \eqref{eq_carbon_dynamics}, carbon mass in atmosphere is affected by previous values of atmosphere and upper ocean carbon mass, using constants parameters $\zeta_{11}$ and $\zeta_{12}$, respectively. Similarly, upper ocean carbon mass is related the past values of carbon mass in atmosphere, upper ocean and lower ocean. Finally, deep ocean carbon mass is not affected by carbon mass in atmosphere and it is affected by past values of ocean related carbon mass, both upper and lower ocean. Also, total emissions $E(i)$ only increases the amount of carbon mass in atmosphere, and according to this model, it takes more than 5 years to see the effect of emissions in carbon mass of upper or lower ocean levels.

\subsection{Climate Dynamics} \label{sec_dice_model_climate}
Temperature deviation of atmosphere, denoted by $T_\mathrm{AT}(i)$, and lower (deep) ocean temperature deviation, denoted by $T_\mathrm{LO}(i)$, are two states to describe the climate model in DICE. These are related to each other and they are driven by the following dynamics:
\begin{equation}
    \begin{bmatrix}
      T_\mathrm{AT}(i+1) \\
      T_\mathrm{LO}(i+1) \\
    \end{bmatrix}
    =
    \begin{bmatrix}
      \phi_{11} & \phi_{12} \\
      \phi_{21} & \phi_{22} \\
    \end{bmatrix}
    \begin{bmatrix}
      T_\mathrm{AT}(i) \\
      T_\mathrm{LO}(i) \\
    \end{bmatrix}
    +
    \begin{bmatrix}
      \xi_1 \\
      0 \\
    \end{bmatrix}
    F(i) \text{,}
    \label{eq_climate_dynamics}
\end{equation}
where elements of state transition and input matrices are constant parameters. The state transition matrix is full, which means both atmosphere and deep ocean temperature deviations, i.e. $T_\mathrm{AT}(i)$ and $T_\mathrm{LO}(i)$, affect each other.

The input to temperature dynamics, $F(i)$, is radiative force resulted by greenhouse effect, which obviously should alter the atmosphere temperature and it is given by:
\begin{equation}
    F(i) = F_{2\times} \log_2 \left( \frac{M_\mathrm{AT}(i)}{M_{\mathrm{AT},1750}} \right) + F_{\mathrm{EX}}(i) \text{,}
    \label{eq_radiative_force}
\end{equation}
where $F_{2\times}$ is the force related to doubling of atmospheric carbon mass, $M_{\mathrm{AT},1750}$ is the carbon mass in atmosphere in year 1750, and $F_{\mathrm{EX}}(i)$ is an exogenous force, described as:
\begin{equation}
    F_{\mathrm{EX}}(i) = f_0 + \min \left\{ \frac{f_1-f_0}{t_f}(i-1), f_1-f_0 \right\} \text{,}
    \label{eq_ex_force}
\end{equation}
with parameters $f_0$, $f_1$ and $t_f$ as minimum force, maximum force, and slope factor, respectively. The starting value of this exogenous force is $f_0$, and it increases by $\frac{1}{t_f}$ every time step, i.e. 5 years, until it reaches the $f_1$. After that, it remains constant.

\subsection{Objective Function of Original DICE Model} \label{sec_dice_model_obj}
Maximization of social welfare function is the original objective of DICE model. The objective function is defined as a discounted sum of utilities within an infinite horizon and it is given by:
\begin{equation}
    W(\mu, s) = \sum_{i=0}^{\infty} \frac{U\big( C(i), L(i) \big)}{(1+\rho)^{i \cdot {\Delta}t}}
    \label{eq_welfare}
\end{equation}
where $\rho$ is the discount rate, $C(i)$ is total consumption, $L(i)$ is total population, and utility function $U(C,L)$ is defined as:
\begin{equation}
    U(C,L) = \left( \frac{\left( \frac{C}{L} \right)^{1-\alpha} - 1}{1 - \alpha} \right) L
    \label{eq_utility_function}
\end{equation}
with $\alpha \geq 0$ as elasticity of marginal utility of consumption. Maximizing the value of $W(\mu,s)$, subject to dynamics of DICE model is a nonlinear optimal control problem.
Because utility function given in \eqref{eq_utility_function} and social welfare defined by \eqref{eq_welfare} are related to consumption, according to \eqref{eq_consumption} and \eqref{eq_net_output}, the objective function is related to both socio-economic factors and climate conditions. However, the function reflects the economic part more than climate-related factors. Hence, in this paper, we are going to introduce another objective, which is directly related to climate.

\subsection{Parameters of DICE Model} \label{sec_dice_model_params}
Parameter values of DICE model are gathered in Table \ref{table_params}. These values are updated in \cite{nordhaus_2016,nordhaus_2017} and used in DICE-2016R version of the model.

\begin{table}[t]
    \caption{Parameters of DICE-2016R Model \cite{nordhaus_2016,nordhaus_2017}}
    \begin{center}
    \begin{tabular}{crll}
    \hline
        \textbf{\thead{Parameter \\ Name}} &
        \textbf{\thead{Value}} &
        \textbf{\thead{Unit}} &
        \textbf{\thead{Equation(s)}} \\
    \hline
    $t_0$                    &  $2015$                         &  -                              &  \eqref{eq_time} \\
    ${\Delta}t$              &  $5$                            &  year                           &  \eqref{eq_time}, \eqref{eq_capital}, \eqref{eq_welfare} \\
    $\ell_g$                 &  $0.134$                        &  -                              &  \eqref{eq_pop_dynamics} \\
    $L_a$                    &  $11,\!500$                       &  million people                 &  \eqref{eq_pop_dynamics} \\
    $\gamma$                 &  $0.3$                          &  -                              &  \eqref{eq_gross_output} \\
    $g_A$                    &  $0.076$                        &  -                              &  \eqref{eq_tech_eff} \\
    $\delta_A$               &  $0.005$                        &  -                              &  \eqref{eq_tech_eff} \\
    $\delta_K$               &  $0.1$                          &  -                              &  \eqref{eq_capital} \\
    $\theta_2$               &  $2.6$                          &  -                              &  \eqref{eq_abatement_cost}, \eqref{eq_mitigation_effort} \\
    $p_b$                    &  $550$                          &  2010-USD/tCO\textsubscript{2}  &  \eqref{eq_mitigation_effort} \\
    $\delta_{pb}$            &  $0.025$                        &  -                              &  \eqref{eq_mitigation_effort} \\
    $\psi_1$                 &  $0$                            &  -                              &  \eqref{eq_climate_damage} \\
    $\psi_2$                 &  $0.00236$                      &  -                              &  \eqref{eq_climate_damage} \\
    $g_\sigma$               &  $0.0152$                       &  -                              &  \eqref{eq_emission_intensity} \\
    $\delta_\sigma$          &  $0.001$                        &  -                              &  \eqref{eq_emission_intensity} \\
    $\delta_\mathrm{EL}$     &  $0.115$                        &  -                              &  \eqref{eq_land_use_emissions} \\
    $E_{\mathrm{L},0}$       &  $2.6$                          &  GtCO\textsubscript{2}/year     &  \eqref{eq_land_use_emissions} \\
    $\zeta_{11}$             &  $0.88$                         &  -                              &  \eqref{eq_carbon_dynamics} \\
    $\zeta_{12}$             &  $0.196$                        &  -                              &  \eqref{eq_carbon_dynamics} \\
    $\zeta_{21}$             &  $0.12$                         &  -                              &  \eqref{eq_carbon_dynamics} \\
    $\zeta_{22}$             &  $0.797$                        &  -                              &  \eqref{eq_carbon_dynamics} \\
    $\zeta_{23}$             &  $0.001465$                     &  -                              &  \eqref{eq_carbon_dynamics} \\
    $\zeta_{32}$             &  $0.007$                        &  -                              &  \eqref{eq_carbon_dynamics} \\
    $\zeta_{33}$             &  $0.99853488$                   &  -                              &  \eqref{eq_carbon_dynamics} \\
    $\xi_1$                  &  $0.1005$                       &  -                              &  \eqref{eq_climate_dynamics} \\
    $\xi_2$                  &  $\frac{3}{11} \approx 0.2727$  &  GtC/GtCO\textsubscript{2}      &  \eqref{eq_carbon_dynamics} \\
    $\phi_{11}$              &  $0.8718$                       &  -                              &  \eqref{eq_climate_dynamics} \\
    $\phi_{12}$              &  $0.0088$                       &  -                              &  \eqref{eq_climate_dynamics} \\
    $\phi_{21}$              &  $0.025$                        &  -                              &  \eqref{eq_climate_dynamics} \\
    $\phi_{22}$              &  $0.975$                        &  -                              &  \eqref{eq_climate_dynamics} \\
    $F_{2\times}$            &  $3.6813$                       &  W/m\textsuperscript{2}         &  \eqref{eq_radiative_force} \\
    $M_{\mathrm{AT}, 1750}$  &  $588$                          &  GtC                            &  \eqref{eq_radiative_force} \\
    $f_0$                    &  $0.5$                          &  W/m\textsuperscript{2}         &  \eqref{eq_ex_force} \\
    $f_1$                    &  $1$                            &  W/m\textsuperscript{2}         &  \eqref{eq_ex_force} \\
    $t_f$                    &  $17$                           &  -                              &  \eqref{eq_ex_force} \\
    $\alpha$                 &  $1.45$                         &  -                              &  \eqref{eq_utility_function} \\
    $\rho$                   &  $0.015$                        &  -                              &  \eqref{eq_welfare} \\
    \hline
    \end{tabular}
    \label{table_params}
    \end{center}

\end{table}

\section{Multi-objective Optimal Control Problem} \label{sec_problem}
In this paper, we are going to modify the original optimal control problem on DICE model into a multi-objective problem. Actually the welfare function defined by \eqref{eq_welfare} contains factors from socioeconomic and climate dynamics, and it reflects the optimality in both directions. However, having an additional straight objective related to climate or carbon mass will provide a complete set of options which can lead to better decisions.

We define a bi-objective optimal control problem, to maximize social welfare and minimize the maximum atmosphere temperature deviation within a specified time window:
\begin{align}
    \max_{\mu, s} &\; W(\mu,s)             \nonumber   \\
    \min_{\mu, s} &\; T_{\mathrm{AT},\max} \label{eq_mo_problem}
\end{align}
subject to dynamics provided by DICE model, i.e. dynamic equations \eqref{eq_pop_dynamics}, \eqref{eq_tech_eff}, \eqref{eq_capital}, \eqref{eq_emission_intensity}, \eqref{eq_carbon_dynamics} and \eqref{eq_climate_dynamics}, and assuming control inputs $0 \leq \mu(i) \leq 1$ and $0 \leq s(i) \leq 1$, for every time index $i$.

Time horizon of this problem is $H$; hence, $i$ varies from $0$ to $H$. Therefor, $T_{\mathrm{AT},\max}$ is defined by following relation:
\begin{equation}
    T_{\mathrm{AT},\max} = \max_{i \in \{ 0, 1, \ldots, H \} } T_{\mathrm{AT}}(i)
    \label{eq_temp_at_max}
\end{equation}
Our second objective is directly related to deviation in temperature and it is a kind of worst-case optimization criterion, very similar to what used in robust control, e.g.  $H_\infty$ control.

Solving the multi-objective control problem, defined by \eqref{eq_mo_problem}, will results in various options, known as Pareto solutions, and this makes possible to make sacrifices in social welfare for the sake of lower temperature deviation and better climate. But single-objective optimal control, misses all of these possibilities and stocks in a specific solution, without any flexibilities.

It is also possible to choose other objectives, for example maximizing the net economic output while minimizing the mass of atmosphere carbon. But according to our tests, the combination of welfare function and maximum temperature deviations, has much better results and provides diverse and meaningful options.

\section{Utilized Algorithm} \label{sec_algorithm}
In this paper, the multi-objective optimal control defined in previous section, is solved using Non-Dominated Sorting Genetic Algorithm (NSGA-II). In this section, we will briefly review the structure of this algorithm and the approached we used to solve the problem.

\subsection{NSGA-II} \label{sec_algorithm_nsga2}
Originally, Srinivas and Deb \cite{srinivas_1994} proposed a new kind of genetic algorithm to deal with multi-objective optimization problems and called it Non-Dominated Sorting Genetic Algorithm (NSGA), which uses the multi-objective domination as ranking (sorting) criterion and fitness sharing mechanism as diversification mechanism. Performance of NSGA is highly sensitive to parameters of fitness sharing procedure. Hence, Deb et al. \cite{deb_2002} introduced a new version of the algorithm, named NSGA-II, which is much faster and less sensitive to parameters. Instead of fitness sharing, NSGA-II uses crowding distance to control the diversity of solutions.

\begin{figure}[t]
    \centerline{\includegraphics[width=\columnwidth]{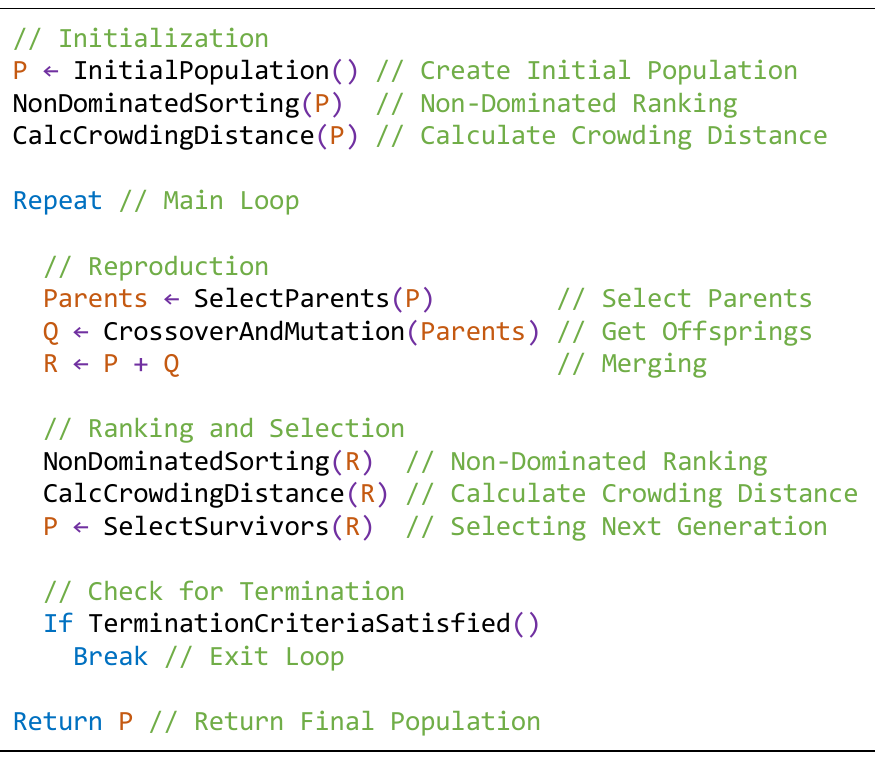}}
    \caption{Pseudo-code of NSGA-II Procedure}
    \label{fig_nsga2}
\end{figure}

After initialization, at every iteration of NSGA-II, the crossover and mutation operations are applied to selected members of current population to create a population of offsprings and mutants Usually, parents are selected using binary tournament selection mechanism in NSGA-II. After creation of new individuals, they are merged into current main population to create a larger population. The resulting population, is sorted according to two criteria: (a) non-domination rank, which is determined by non-dominated sorting procedure, and (b) crowding distance, which controls the diversification by removing similar solutions and keeping distinct ones. Finally, members of the population with rank 1 (first front, F\textsubscript{1}), will form Pareto frontier. A pseudo-code of NSGA-II is given in Fig. \ref{fig_nsga2}.

Solutions returned by NSGA-II are not dominated by other solutions, i.e. they are first-rank members of the population. These solutions form Pareto front and they can be seen as multi-objective optimal, or at least sub-optimal, solutions of the problem being solved. Decision maker can select any of these solutions as final strategy, while being sure about optimality of the final outcome.

\subsection{Solution Structure and Initialization} \label{sec_algorithm_solution}
According to \eqref{eq_mo_problem} and \eqref{eq_temp_at_max}, we have two control inputs and we must determine the value of these signals at every time step. So the total number of decision variables is $2 \times H$, and all of them are in the range $[0,1]$. The structure of the solutions is as follows:
\begin{equation}
    \begin{bmatrix}
      \mu(0) & \mu(1) & \cdots & \mu(i) & \cdots & \mu(H-1) \\
      s(0)   & s(1)   & \cdots & s(i)   & \cdots & s(H-1)   \\
    \end{bmatrix}
    \in [0,1]^{2 \times H} .
    \label{eq_solution_structure}
\end{equation}

Note that input at time step $H$ will not have any effects on the states, which are used to calculate the objective functions. Actually, they affect the states at time step $H+1$, which is out of problem time window.

In the initialization phase of the algorithm, solutions are created assuming constant inputs over time, i.e. all elements in first row of solution matrix have same value, as well as the elements of second row, sharing their own common value. Creating initial solutions according to this scheme, results in a better Pareto frontier.

\section{Results} \label{sec_results}
Dynamics of DICE model, objectives, and NSGA-II are implemented using MATLAB. Parameters of the algorithm are set as follows:
\begin{itemize}
  \item Maximum Number of Iterations: 1000,
  \item Population Size: 200,
  \item Mutation Rate: 3\%,
  \item Mutation Step Size: 0.1, and
  \item Mutation Type: Gaussian (Normal),
\end{itemize}
and the time horizon ($H$) is set to be 37. So we seek for Pareto-optimal values of decision variables in a 74-dimensional search space. The time window of our simulation of DICE model, is from year 2015 to 2200.

\begin{figure}
    \centerline{\includegraphics[width=\columnwidth]{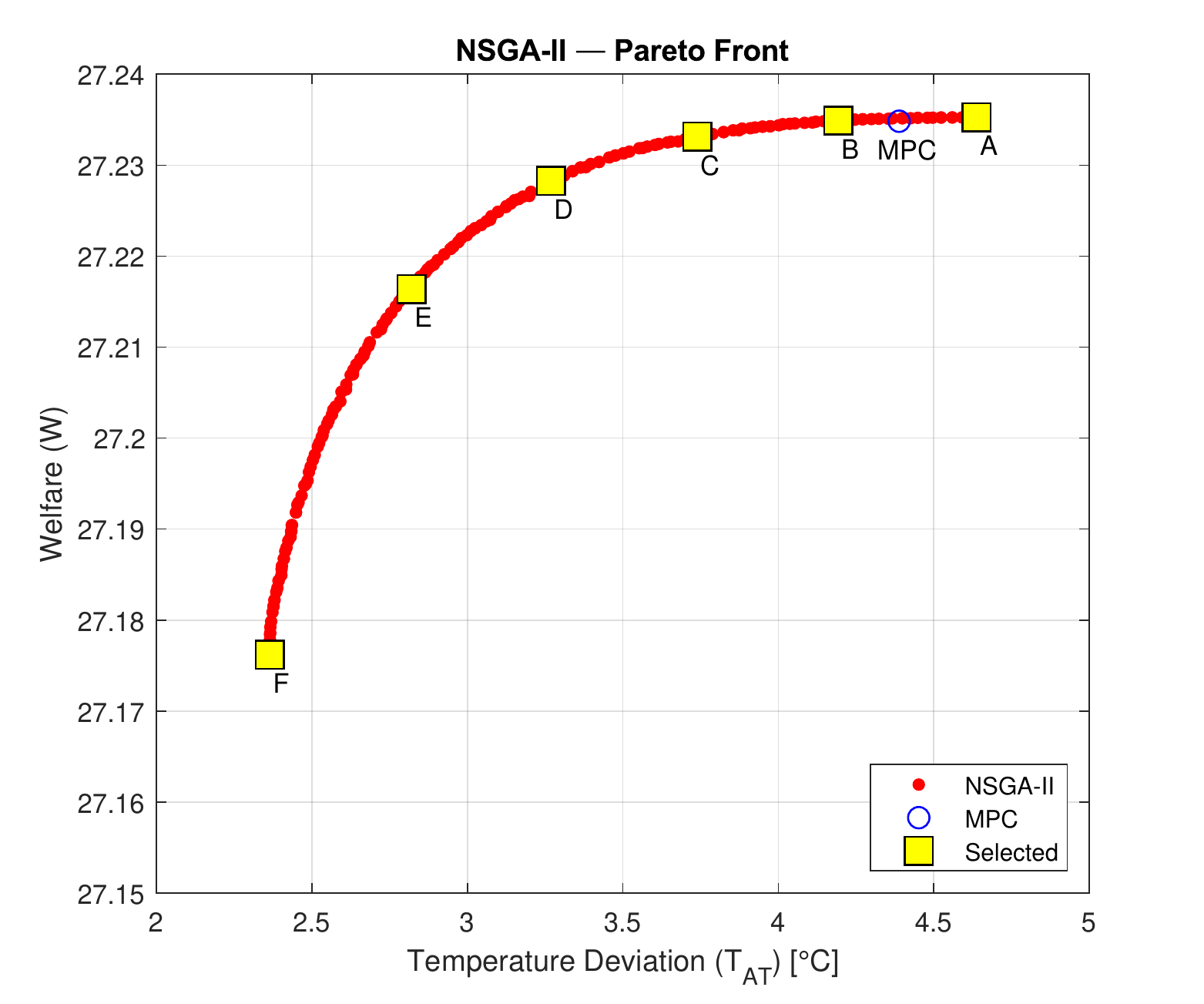}}
    \caption{Resulting Pareto frontier, also solution of MPC}
    \label{fig_pareto_front}
\end{figure}

Fig. \ref{fig_pareto_front} illustrates the resulting Pareto frontier. To compare the results with \cite{faulwasser_2018} and \cite{kellett_2019}, the solution of Model Predictive Control (MPC) is shown in bi-objective plane, as well. According to this plot, result of MPC is very close to be Pareto-optimal. However, the solution provided by MPC is dominated by some points on the Pareto frontier and it is not a Pareto-optimal solution. Having this Pareto front, makes possible to make better decisions, for example compromise about reducing in socioeconomic objective of welfare, to have better climate.

Some of solutions, marked with A to F, are selected from Pareto set, with equal spacing with respect to temperature deviation (first objective). Solution A is related to maximum social welfare (best) and maximum temperature deviation (worst). Similarly, solution F is related to minimum deviation of atmosphere temperature (best) and minimum welfare (worst). According to path from F to A, increasing social welfare will result the temperature deviation, and this is because of conflicting nature of objective functions.

As mentioned earlier in subsection \ref{sec_dice_model_obj}, social welfare function is related to both socio-economic and climate-related factors. However, it is mainly affected by economic variables, such as gross economic output and it is less sensitive to environment conditions. This can be inferred from Fig. \ref{fig_pareto_front}, too. The result of MPC is very close to the upper-left part of the Pareto frontier. The single-objective approach, which is aimed to maximize the social welfare only, discovers a solution with almost highest possible welfare and temperature deviation. This show the true nature of social welfare function and its sensitivity to climate-related factors, such as atmosphere temperature deviation.

\begin{figure}
    \centerline{\includegraphics[width=\columnwidth]{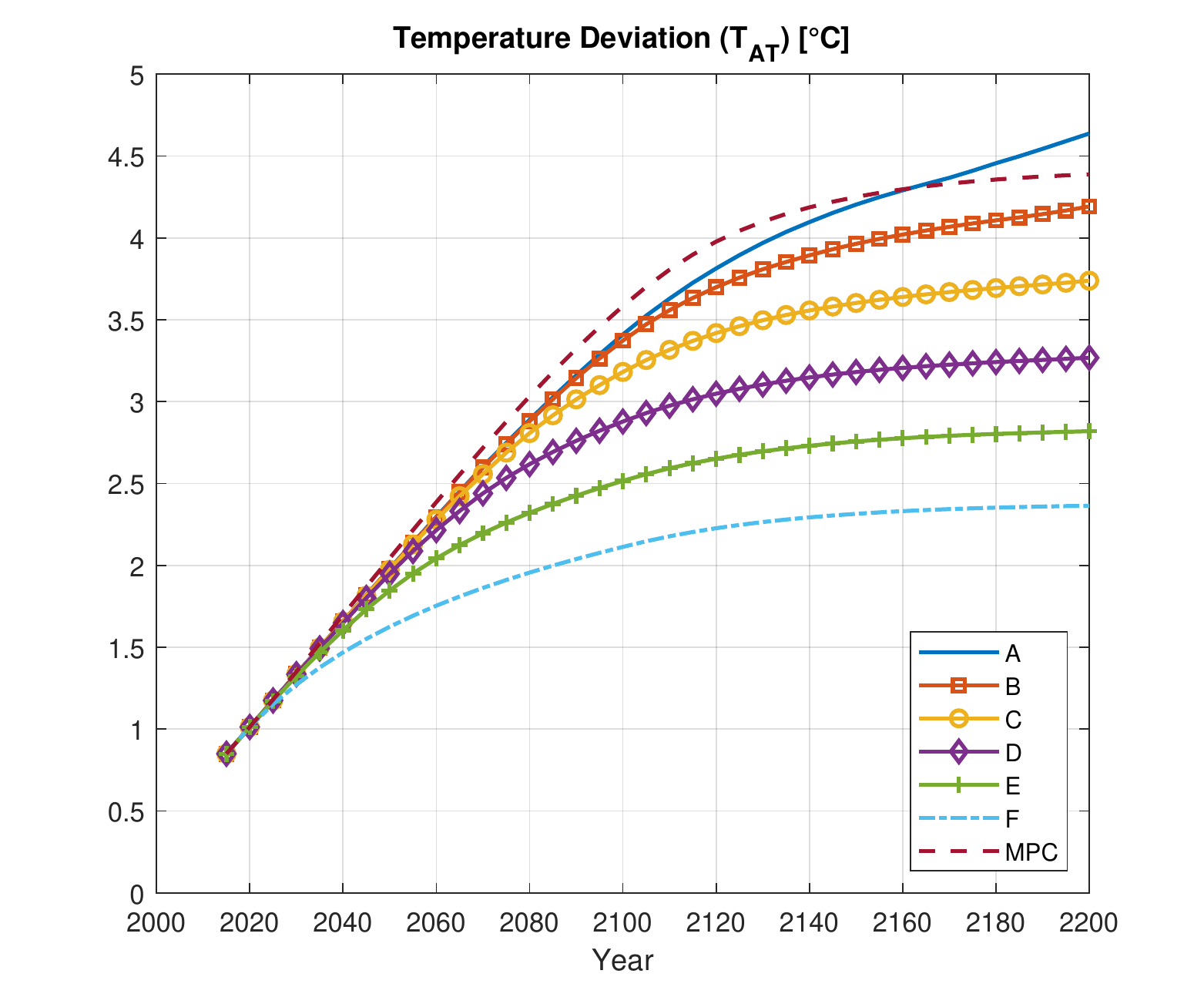}}
    \caption{Atmosphere temperature deviation $T_\mathrm{AT}$ for selected Pareto-optimal solutions, A to F, and MPC result}
    \label{fig_temp_dev}
\end{figure}

\begin{figure}
    \centerline{\includegraphics[width=\columnwidth]{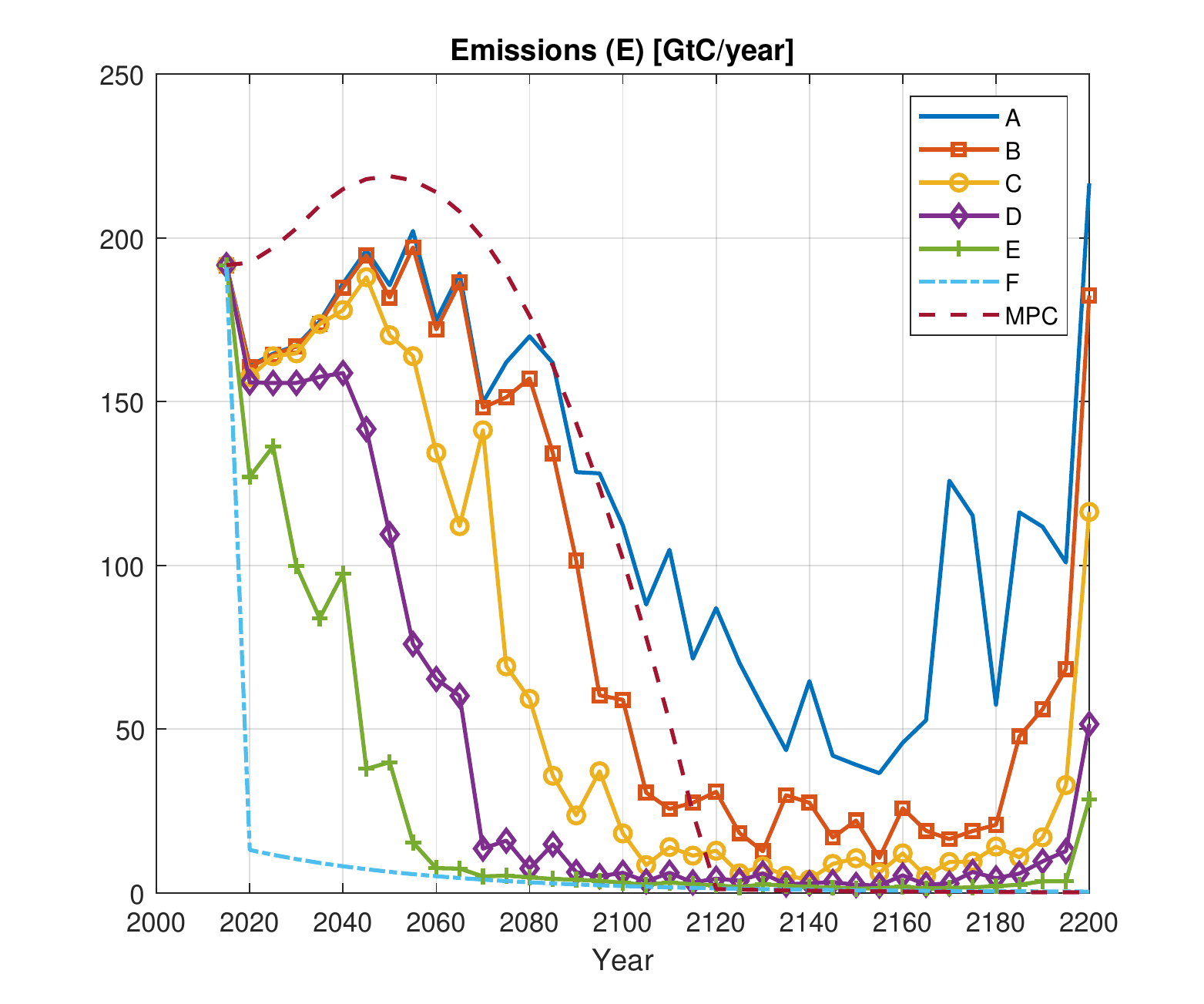}}
    \caption{Total emissions $E$ for selected Pareto solutions and MPC results}
    \label{fig_emissions}
\end{figure}

\begin{figure}
    \centerline{\includegraphics[width=\columnwidth]{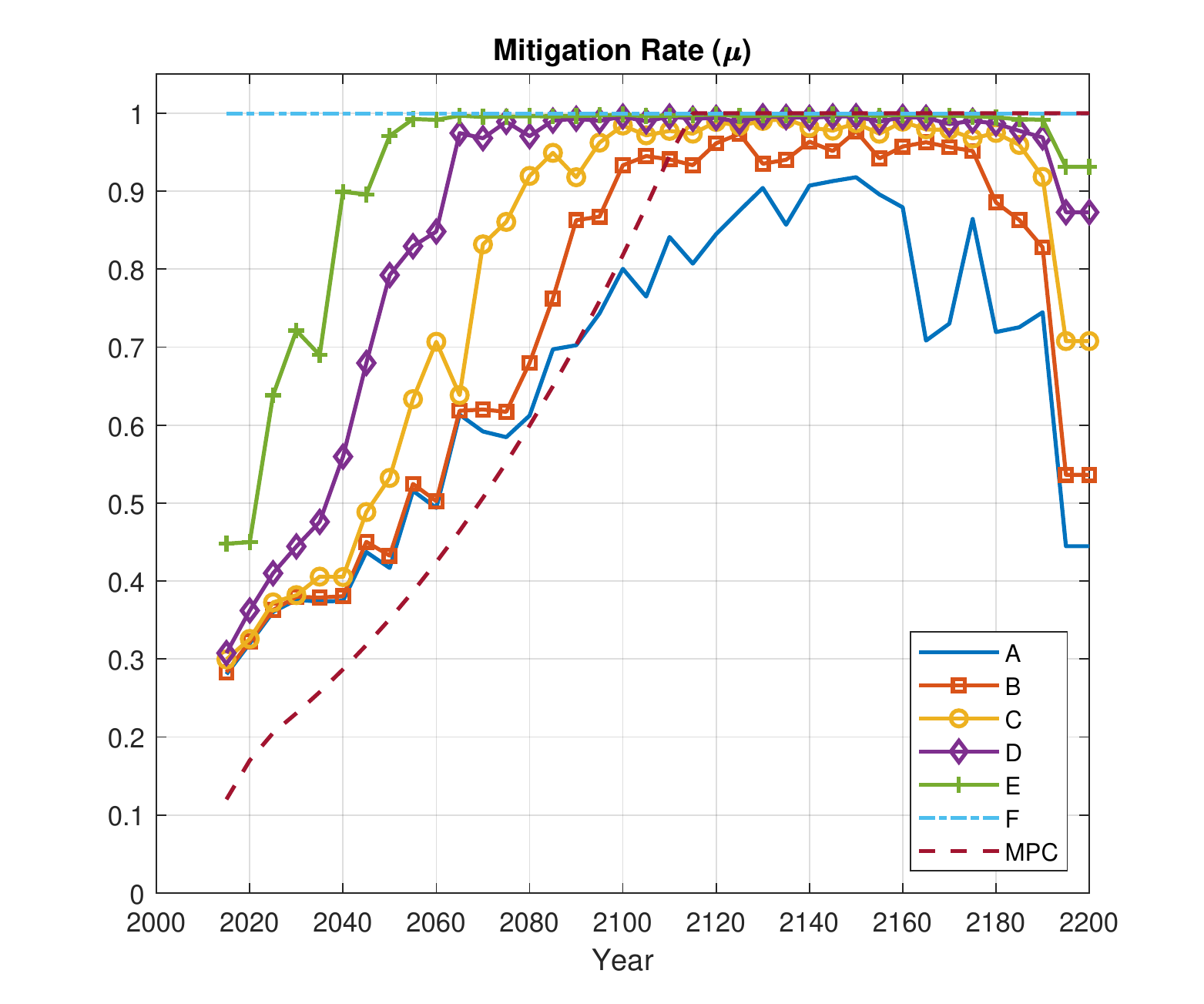}}
    \caption{Mitigation rate $\mu$ control input for selected Pareto-optimal solutions, A to F, and MPC result}
    \label{fig_mitigation_rate}
\end{figure}

\begin{figure}
    \centerline{\includegraphics[width=\columnwidth]{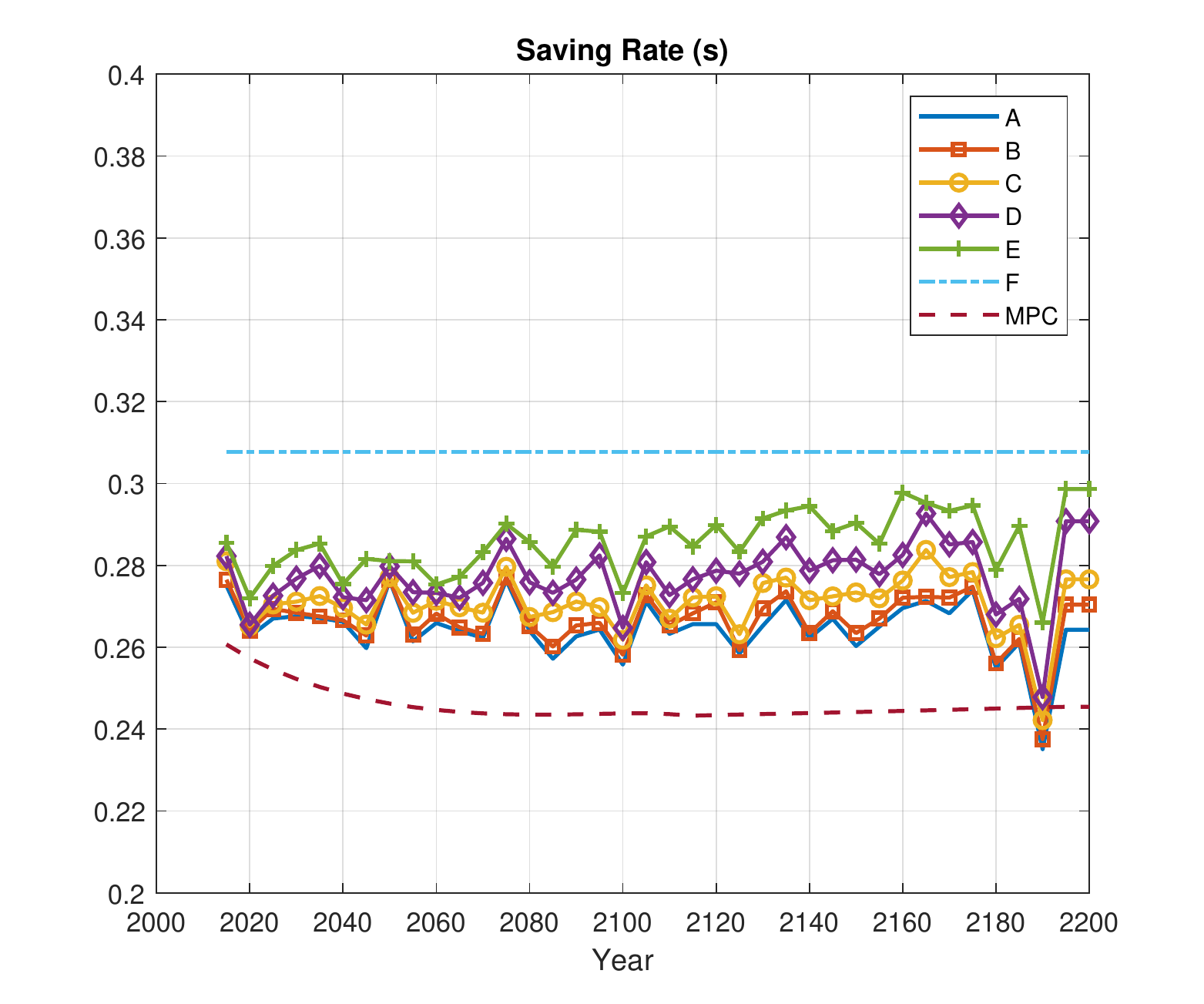}}
    \caption{Saving rate $s$ input control for selected Pareto-optimal solutions, A to F, and MPC result}
    \label{fig_saving_rate}
\end{figure}

Some of states and control inputs for these selected solutions are shown in Fig. \ref{fig_temp_dev} to Fig. \ref{fig_saving_rate}. These plots illustrate the deviation of atmosphere temperature (from temperature in year 1900) $T_\mathrm{AT}$, total emissions $E$, mitigation rate $\mu$ (first control input), and saving rate s (second control input), respectively. In addition to selected solutions from Pareto set, the same signals are plotted for MPC results \cite{faulwasser_2018,kellett_2019}, too.

Objective values of selected solutions, A to F, and solution found by MPC are listed in Table II. The rows of table are sorted in descending order of social welfare value. According to the objective values, for example, moving from MPC to D, will reduce the welfare by 0.0077 (less than 0.03\%), but it will reduce the temperature deviation by more than 1.15 $^\circ$C (more than 26 percent, relatively).

However, the results indicate that it is not possible to have temperature deviations less than 2.37 $^\circ$C, with current technologies and efforts. In best situation, this is the minimum reachable temperature, if everything goes as predicted by DICE model. Unless, some technological advancements will emerge, and makes it possible to go below this limit.

\begin{table}[t]
    \caption{Objective Values for Selected Solutions and MPC}
    \begin{center}
    \begin{tabular}{crrrr}
    \hline \\[-10pt]
        \multirow{2}{*}{\textbf{\makecell{Solution \\ Name}}} &
        \multirow{2}{*}{\textbf{\makecell{Social \\ Welfare, \\ $W$}}} &
        \multirow{2}{*}{\textbf{\makecell{Maximum \\ Temperature \\ Deviation, \\ $T_{\mathrm{AT},\max}$ [$^\circ$C]}}} &
        \multicolumn{2}{c}{\textbf{\makecell{Changes Relative to \\ MPC Objective Values}}} \\[10pt]
        \cline{4-5} \\[-10pt]
        &&& $\Delta W$ & $\Delta T_\mathrm{AT}$ \\[4pt]
    \hline
    A    &  $27.2360$  &  $4.5380$  &  $ 0.0012$  &  $ 0.1495$  \\
    MPC  &  $27.2348$  &  $4.3885$  &  $ 0     $  &  $ 0     $  \\
    B    &  $27.2354$  &  $4.1100$  &  $ 0.0006$  &  $-0.2785$  \\
    C    &  $27.2332$  &  $3.6862$  &  $-0.0016$  &  $-0.7023$  \\
    D    &  $27.2271$  &  $3.2368$  &  $-0.0077$  &  $-1.1517$  \\
    E    &  $27.2147$  &  $2.8066$  &  $-0.0201$  &  $-1.5819$  \\
    F    &  $27.1600$  &  $2.3768$  &  $-0.0748$  &  $-2.0117$  \\
    \hline
    \end{tabular}
    \label{table_results}
    \end{center}

\end{table}

\section{Conclusion} \label{sec_conclusion}
A bi-objective optimal control problem is defined on DICE model, to find Pareto-optimal strategies for carbon mitigation (abatement) and saving rate (investment), which maximize the social welfare function while minimizing the deviation of atmosphere temperature. This problem is solved by NSGA-II, a multi-objective genetic algorithm and a Pareto frontier is found. There are solutions in this set which dominate the result of previous works on model predictive control of DICE model. Also, the Pareto front offers a wide range of options to reach lower temperature deviation by sacrificing welfare. However, the results show that, unfortunately, it is not possible to go below some limits and decrease the temperature deviation as much as desired. Application of the same technique on RICE model, a regional version of DICE which is also developed by Nordhaus, and changing the structure of algorithm to find smooth control inputs, can be possible directions for future research in this field.

\bibliographystyle{ieeetr}
\bibliography{refs}
\balance

\end{document}